\documentclass[conference,a4paper]{IEEEtran}
\usepackage{cite}
\ifCLASSINFOpdf
  \usepackage[pdftex]{graphicx}
  \graphicspath{{../pdf/}{../jpeg/}}
   \DeclareGraphicsExtensions{.pdf,.jpeg,.png}
\else
\fi
\usepackage[cmex10]{amsmath}
\usepackage{xfrac}
\usepackage{array}
\usepackage{mdwmath}
\usepackage{amsfonts}
\usepackage{amssymb}
\usepackage{amsbsy}
\usepackage{amsthm}
\usepackage{tikz}
\usetikzlibrary{external}
\begin{document}
\title{Analysis of Coverage Region for MIMO Relay Network with Multiple Cooperative DF-Relays}




\author{\IEEEauthorblockN{Behrooz Razeghi,
Alireza Alizadeh,
Sima Naseri,
Ghosheh Abed Hodtani,
Seyed Alireza Seyedin}
Department of Electrical Engineering, Faculty of Engineering\\
Ferdowsi University of Mashhad,
Mashhad, Iran\\
E-mail: behrooz.razeghi.r@ieee.org, \{alr.alizadeh, naseri.sima1990, ghodtani\}@gmail.com, seyedin@um.ac.ir
%E-mail: behrooz.razeghi.r@ieee.org, alr.alizadeh@gmail.com, naseri.sima1990@gmail.com,\\ ghodtani@gmail.com, seyedin@um.ac.ir
 }

\IEEEoverridecommandlockouts
\IEEEpubid{\makebox[\columnwidth]{978-1-4799-5863-4/14/\$31.00~\copyright2014
IEEE \hfill} \hspace{\columnsep}\makebox[\columnwidth]{ }}

\maketitle

\begin{abstract}
%\boldmath
We study and analyze coverage region in MIMO communication systems for a multiple-relay network with decode-and-forward (DF) strategy at the relays. Assuming that there is a line-of-sight (LOS) propagation environment for source-relay channels and channel state information is available at receivers (CSIR), we consider the objective of maximizing coverage region for a given transmission rate and show numerically the significant effect of propagation environment on capacity bounds, optimal relay location and coverage region. Also, we study the situation in which two adjacent relays cooperate in transmission signals to the destination and show analytically that the coverage region is extended compared to noncooperative scenario.
 
\end{abstract}

%  keywords

% \fi
\IEEEpeerreviewmaketitle

\section{Introduction}
%no \IEEEPARstart
The relay channel is the most basic structural unit in wireless networks and relaying can increase coverage region and transmission rate between the source and the destination. Relaying strategy can realize some of the gains of multiple-antenna systems by single-antenna terminals, i.e., the relay nodes act as a distributed multi-antenna system. 

The relay channel, first introduced by Van der Meulen \cite{van1971three}, was studied in detail by Cover-El Gamal in \cite{cover1979capacity}.
In \cite{hodtani2009unified} known capacity theorems for the relay channel have been unified into one capacity theorem.

In \cite{kramer2005cooperative} Kramer \emph{et al.} examined Gaussian relay channel and considered the effect of relocating the relay on achievable rates at the destination. However, in many practical cases the location of the relay is determined at the time of network design and the design problem is to maximize coverage region for a desired transmission rate. In \cite{aggarwal2009maximizing} the authors studied the problem of maximizing coverage for a given rate and evaluated decode-and-forward (DF) and compress-and-forward (CF) strategies with the objective of maximizing coverage for Gaussian point to point relay channel. The authors of \cite{zhao2007coverage} analyzed the coverage extension by using decode-and-forward (DF) relays in a cellular system and found coverage range for two special cases corresponding to upper and lower bounds of deterministic MIMO relay channel capacity. 

In \cite{alizadeh2012analysis} and \cite{alizadeh2013analysis} the authors analyzed the coverage region in MIMO relay channel with single relay in Rayleigh fading case and determined the optimal relay location maximizing the coverage region. The authors of \cite{Fanny2014impactfading} studied coverage region and energy efficiency for Gaussian relay channel for a specific network geometry. 
%The coverage region of MIMO relay channel with amplify-and-forward relaying strategy has been studied in \cite{razeghi2014AFdirectlink}.

%%%  our work
\textbf{Our work}: 
In this paper, a MIMO relay network with multiple relays is considered in which, we (1) extend the obtained results for coverage region in \cite{Fanny2014impactfading} to multi-antenna communication system, (2) generalized the single relay channel studied in \cite{alizadeh2012analysis} and \cite{alizadeh2013analysis} to MIMO relay network. Also, we investigate the effect of channel fading on the capacity bounds, optimal relay location, and coverage region. Next, we show analytically that the circular coverage region is extended in relays' cooperation scheme. Since our goal is to obtain maximum coverage region, we put relays at maximum distance for a certain rate at the relay such that if we increase the distance, the relay cannot decode satisfactorily. We require both relay and destination to fully decode the received signals. The relay has sufficient power supply and better antennas than the destination. Furthermore, since the relay is generally placed in LOS scenario of the source, the source-relay channel is stronger than the relay-destination and source-destination channels. More precisely, for the border users which have weaker channel, the decode-and-forward transmission rate is limited to the source-destination transmission rate. Thus, to determine coverage region we should assume successful decoding at both relay and destination.
%%%%%%%%%%%%% %
\begin{figure}[t]
\centering
\includegraphics[scale=1]{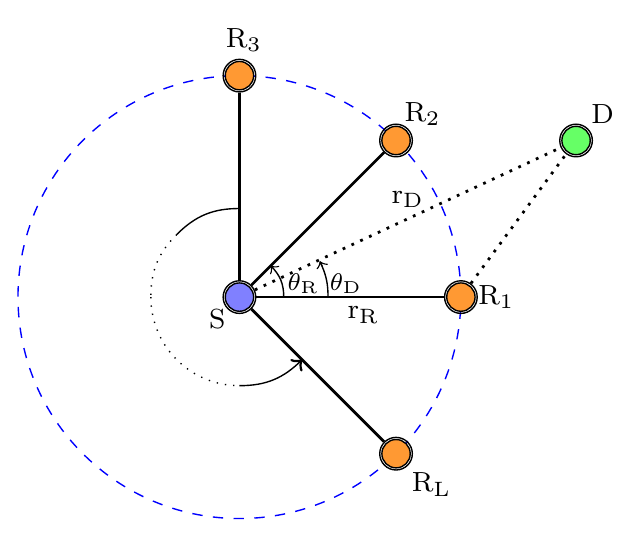}
    \caption{Network geometry.}
    \label{Fig1}
\end{figure}

\textbf{Notations}: Throughout this paper, we use $\mathbb{E}\left\{.\right\}$ to denote the expectation operator; "$\dag$" stands for the conjugate transpose; the distribution of a circularly symmetric complex Gaussian vector with mean $\mathbf{m}$ and covariance matrix $\mathbf{Q}$ is denoted as $\mathcal{CN}(\mathbf{m},\mathbf{Q})$; and vectors and matrices are denoted by boldface lower case ($\mathbf{x}$) and upper case letters ($\mathbf{X}$), respectively.

The rest of the paper is organized as follows. In Section II, we present channel model and define the concept of coverage region. We review in Section III the capacity bounds of MIMO relay channel. In Section IV we recapitulate the main results of desired transmission rate in the terms of optimal relay location. We then generalize our scheme to cooperative scenario in Section V. Simulation results are presented in Section VI, and Section VII contains our conclusion.

%I wish you the best of success.


%\hfill January 11, 2007
 
%%%%%%%%%%%%%%%%%%% 
\section{Channel Model and Preliminaries}
\subsection{Channel Model}
\begin{figure}[t]
\centering
\includegraphics[scale=0.94]{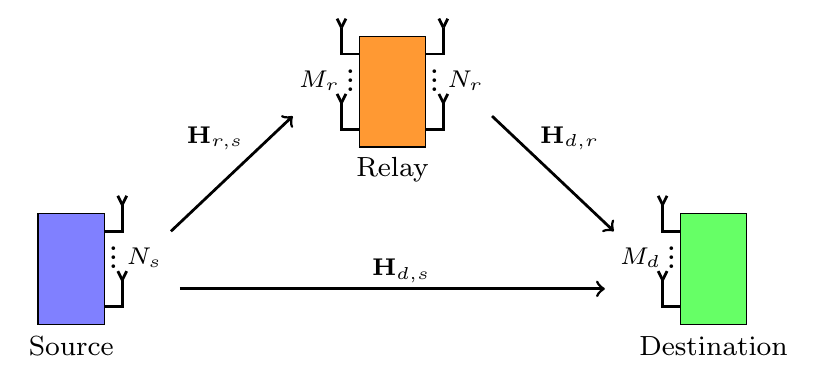}
    \caption{MIMO relay channel.}
    \label{Fig2}
\end{figure}

%\subsection{\textbf{Channel Model}}
Consider one source, $L$ relays , and one destination. The relays are placed uniformly on a circle around the source and they divide the cell into equal sized sectors. Our network geometry is depicted in Fig.~\ref{Fig1}. In our model, the relay node is assumed to be full-duplex. 
We assume each relay supports all users in its sector, and different sub-carrier is assigned to each relay. Our MIMO relay channel of each sector is depicted in Fig.~\ref{Fig2}. 

The received signals at the relay node and destination node in DF scenario can be written as (the sector index is dropped for simplicity)
\vspace{-5pt}
\begin{align}\label{df.model}
     \mathbf{y}_r &= \mathbf{H}_{r,s}\mathbf{x}_s + \textbf{z}_r \nonumber \\
     \mathbf{y}_d &= \mathbf{H}_{d,s}\mathbf{x}_s + \mathbf{H}_{d,r}\mathbf{x}_r + \textbf{z}_d
\end{align}
\vspace{-2pt}
where
\begin{itemize}
  \item $\mathbf{x}_s,\,\mathbf{x}_r$ are $N_s \times 1$ and $N_r \times 1$ transmitted signals from the source and relay, respectively.
      The power constraints are $\mathbb{E}\left\{ \mathbf{x}_s^{\dag} \mathbf{x}_s \right\} \leq P_s$
      and $ \mathbb{E}\left\{ \mathbf{x}_r^\dagger \mathbf{x}_r \right\}\leq P_r$.
  \item $\mathbf{y}_r,\,\mathbf{y}_d$ are $M_r \times 1$ and $M_d \times 1$ received signals at the relay and destination, respectively.
  \item $\mathbf{H}_{r,s},\,\mathbf{H}_{d,s}$,\, and $\mathbf{H}_{d,r}$,\, are $M_r \times N_s,\, M_d \times N_s$, and $M_d \times N_r$ channel gain matrices.
  \item $\textbf{z}_r,\,\textbf{z}_d$ are independent $M_r \times 1$ and $M_d \times 1$ circularly symmetric complex Gaussian noise vectors with distributions $\mathcal{CN}(\mathbf{0},\mathbf{I}_{M_r})$ and $\mathcal{CN}(\mathbf{0},\mathbf{I}_{M_d})$.
\end{itemize}

Considering $N=N_s+N_r$, the $N\times N$ covariance matrix of the input signals can be written as
\begin{equation}
\label{joint.cov}
\mathbf{Q}\triangleq \mathbb{E} \left\{ {\left[\mathbf{x}_s \; \mathbf{x}_r\right]}^t  {\left[\mathbf{x}_s \; \mathbf{x}_r\right]}^\dag \right\} =
\begin{bmatrix}
\mathbf{Q_{ss}}& \mathbf{Q_{sr}}\\
\mathbf{Q_{rs}}& \mathbf{Q_{rr}}
\end{bmatrix}
\end{equation}
where $\mathbf{Q}_{ij}=\mathbb{E}\left\{\mathbf{x}_i \mathbf{x}_j^{\dag}\right\}, i,j=s,r$ is the covariance matrix between the input signals $\mathbf{x}_i$ and $\mathbf{x}_j$. Note that $\mathbf{Q}$ is a Hermitian matrix.

The source-relay channel gain matrix can be modeled as \cite{paulraj2003introduction}
% oestges2004propagation
\begin{align}\label{sr.H}
    \mathbf{H}_{r,s} &= \sqrt{P_{r,s}}\mathbf{\widetilde{H}}_{r,s} \nonumber \\ &= \sqrt{P_{r,s}}\left(\sqrt{\frac{K}{K+1}}\mathbf{\widetilde{H}}_{r,s}^{\mathrm{LOS}} + \sqrt{\frac{1}{K+1}}\mathbf{\widetilde{H}}_{r,s}^{\mathrm{NLOS}}\right)
\end{align}
where
$\sqrt{\sfrac{K}{K+1}}\mathbf{\widetilde{H}}_{r,s}^{\mathrm{LOS}}$ is line-of-sight (fixed) component of the channel. $\sqrt{\sfrac{1}{K+1}}\mathbf{\widetilde{H}}_{r,s}^{\mathrm{NLOS}}$ is non line-of-sight (variable) component that assumes uncorrelated fading and takes into account the influence of the scattering components during the propagation and a tilde on the head denotes the normalized channel matrices. $K$ is the Rician factor for source-relay channel and is the ratio of the total power in the fixed component of the channel to the power in fading component. In \cite{3GPP2005} Rician $K$-factor is modeled as a function of distance between transmitter and receiver. 

The channel matrix for LOS component can be modeled as \cite{jiang2007modelling}
%\vspace{-3}
\begin{align}\label{los.h}
    \mathbf{H}^{\textrm{LOS}} &=
    \begin{bmatrix}
        \mathbf{\acute H}_{M_rN_s} \odot \mathbf{A}_{M_rN_s}^{VV} & \mathbf{\acute H}_{M_rN_s} \odot \mathbf{A}_{M_rN_s}^{VH}\\
        \mathbf{\acute H}_{M_rN_s} \odot \mathbf{A}_{M_rN_s}^{HV} & \mathbf{\acute H}_{M_rN_s} \odot \mathbf{A}_{M_rN_s}^{HH}
    \end{bmatrix}
\end{align}
where $\odot$ denotes the element wise multiplication. $\mathbf{\acute H}_{M_rN_s}$ is LOS source-relay channel matrix with co-polarized antennas. $\mathbf{A}_{M_rN_s}$ is the matrix representing the polarization mismatch. We assume that transmit and receive antennas at the source node and the relay nodes are all strictly aligned; this means we do not take into account the polarization antennas. In this condition, $\mathbf{A}_{M_rN_s}^{VV}$ and $\mathbf{A}_{M_rN_s}^{HH}$ are all-one matrices, while $\mathbf{A}_{M_rN_s}^{VH}$ and $\mathbf{A}_{M_rN_s}^{HV}$ are all-zero matrices. We take into account two alternative prototypes of $\mathbf{\acute H}_{M_rN_s}$ which correspond to poorly-conditioned channel and well-conditioned channel in LOS scenario.

\subsection{Coverage Region}
%\subsection{\textbf{Coverage Region}}

We assume isotropic channel conditions. Thus, coverage region in the absence of relays is circular. We use the meaning of coverage range and coverage angle as defined in \cite{zhao2007coverage}. The coverage angle is defined as $\theta_{cov} = \left(\sfrac{360^\circ}{L}\right)$. This means that $L$ relays are placed uniformly on a circle surrounding the source and they divide the cell into equal sized sectors with angle $\theta_{cov}$. The coverage range is defined as the maximum radius of circular area achieved by placing those $L$ relays.
%%%%%%%%%%%%%
Now, consider the network geometry depicted in Fig.~\ref{Fig1}. In this configuration the source node is located at $d_S = \left(0,0\right)$ and the relay nodes are located at $d_{R_n} = \left({r_R}, (n-1)\theta_R\right), \, n = 1, ..., L$ which $|\theta_R|=\theta_{cov}$. %for $n=1,...,N_r$.
The destination node is located at $d_D = \left(r_D, \theta_D\right)$. We let $\alpha$ be the path loss component. Thus, the channel gain matrices can be written as
\begin{align}\label{pathloss.H}
    \mathbf{H}_{r,s} &= \frac{1}{r_R^{\sfrac{\alpha}{2}}}\mathbf{\hat{H}}_{r,s}, \: \:  \:  \: \;
    \mathbf{H}_{d,s} = \frac{1}{r_D^{\sfrac{\alpha}{2}}}\mathbf{\hat{H}}_{d,s}, \nonumber \\
    \mathbf{H}_{d,r} &= \frac{1}{\left({r_D^2 + r_R^2 - 2 r_D r_R \cos \phi}\right)^{\sfrac{\alpha}{2}}} \mathbf{\hat{H}}_{d,r}, \;
\end{align}
where the entries of $\mathbf{\hat{H}}_{r,s}$, $\mathbf{\hat{H}}_{d,s}$, and $\mathbf{\hat{H}}_{d,r}$ are i.i.d. $\mathcal{CN}(0,1)$, and consequently, $\mathbf{\hat{H}}_{r,s} \mathbf{\hat{H}}^{\dagger}_{r,s}$, $\mathbf{\hat{H}}_{d,s} \mathbf{\hat{H}}^{\dagger}_{d,s}$, and $\mathbf{\hat{H}}_{d,r} \mathbf{\hat{H}}^{\dagger}_{d,r}$ are central complex Wishart matrices with
identity covariance matrix \cite{tulino2004random}. Furthermore, we assume that each block's use of the channel corresponds to an independent realization of channel matrices. The variable $\phi$ is angle between the destination and its corresponding relay. We let $r_{DR}=\left( {r_D^2 + r_R^2 - 2 r_D r_R \cos \phi}\right)$ denotes relay-destination distance.

Analogous to \cite{aggarwal2009maximizing}, now we define the concept of coverage as
%\vspace{-5}
\begin{align}\label{def.coverage}
    \mathcal{C}\! \mathit{o}\mathit{v} \left(r_R\right) = \left\{d_D : C\left(r_R,d_D\right) \geq R_c \right\}
\end{align}
where $R_c > 0$ denotes the desired transmission rate and $C$ denotes the capacity at the relay location $r_R$ and destination node location $d_D$.
%%%%%%%%%%%%%%%%%%%%%%%%%%%%%%%%%%%%%%%

\section{Capacity Bounds for MIMO Relay Channel}
In this section we review the capacity upper bound and lower bound for MIMO relay channel.
\subsection{Cut-set Upper Bound}
%\subsection*{\textbf{Cut-set Upper Bound}}

The capacity upper bound of the discrete memoryless relay channel is
\begin{align}\label{cupper}
C_{\mathrm{upper}}\! =\! \! \mathop{\max}_{p(x_s,x_r)}\! \! \min \left\{I\! \left(X_s,X_r;Y_d\right)\! , I\! \left(X_s;Y_r,Y_d \mid \! X_r\right)\right\}
\end{align}
where the first term under the minimum corresponds to cooperative multiple-access (MAC) bound, the second one corresponds to cooperative broadcast (BC) bound, and the maximization is with respect to the joint distribution of the source and relay signals. Considering ${\mathbf{x}_i\sim\mathcal{CN}(\mathbf{0},\mathbf{Q}_{ii}), i=s,r}$, where $\mathbf{Q}_{ii}$ is the covariance matrix of $\mathbf{x}_i$, the mutual information expressions in~\eqref{cupper} can be expressed for the MIMO relay channel as~\cite{foschini2011opt}  %%\cite{wang2005capacity, foschini2011opt}
\begin{align}
\label{Ccs}
C_{\mathrm{CS}} = \max_{\mathbf{Q}_{ii}:\mathrm{tr}(\mathbf{Q}_{ii})\leq P_i,~i=s,r} \min{(C_1,C_2)}
\end{align}
%	\vspace{-7}
\begin{align}
\label{c1}
C_1 &= \log \det(\mathbf{I}_M+\mathbf{H}_{\textmd{BC}}\mathbf{Q}_{s|r}\mathbf{H}_{\textmd{BC}}^\dag)
\end{align}
\begin{align}
\label{c2}
C_2 &= \log \det(\mathbf{I}_{M_d}+\mathbf{H}_{\textmd{MAC}}\mathbf{Q}\mathbf{H}_{\textmd{MAC}}^\dag)
\end{align}
where $\mathbf{H}_{\textmd{BC}}=\begin{bmatrix}\mathbf{H}_{d,s}\\\mathbf{H}_{r,s}\end{bmatrix}$, $\mathbf{H}_{\textmd{MAC}} = \begin{bmatrix}\mathbf{H}_{d,s}&\mathbf{H}_{d,r}\end{bmatrix}$, $M=M_r+M_d$
and $\mathbf{Q}_{s|r} \triangleq \mathbb{E}\left\{\mathbf{x}_s\mathbf{x}_s^{\dag}|\mathbf{x}_r\right\}=
\mathbf{Q}_{ss}-\mathbf{Q}_{sr}\mathbf{Q}_{rr}^{-1}\mathbf{Q}_{rs}$ is the conditional covariance matrix and given by Schur complement of $\mathbf{Q}_{rr}$ in $\mathbf{Q}$ ~\cite{boyd2004convex}.
The optimal distribution $p(x_s,x_r)$ in~\eqref{cupper} for Gaussian relay channel is Gaussian~\cite{cover1979capacity}, and consequently the maximization of~\eqref{Ccs} would be with respect to three covariance matrices $\mathbf{Q}_{ss}$, $\mathbf{Q}_{rr}$, and $\mathbf{Q}_{sr}$.

When the channel matrices are random and the CSI is only known at the receivers, the optimal joint transmit covariance matrix $\mathbf{Q}$ in~\eqref{Ccs} is diagonal. The authors in \cite{infotheoricrelay} showed that the equal power allocation is the optimal solution, i.e.,
\begin{equation}
\label{cov.matrix}
\mathbf{Q}_{ss}=\frac{P_s}{N_s}\mathbf{I}_{N_s},~~\mathbf{Q}_{rr}=\frac{P_r}{N_r}\mathbf{I}_{N_r},~~
\mathbf{Q}_{sr}=\mathbf{0}
\end{equation}
where $\mathbf{Q}_{sr}=\mathbf{0}$ refers to the independence between the source and the relay signals. Thus, the CS upper bound for the MIMO relay channel with only CSIR can be expressed as
\begin{equation}
C_{\mathrm{CS}}=\min(C_{1},C_2)
\end{equation}
\begin{equation}
\label{C1}
C_1= \mathbb{E} \left\{\log\det(\mathbf{I}_{M}+\frac{P_s}{N_s}\mathbf{H}_\textmd{BC} \mathbf{H}_{\textmd{BC}}^\dag)\right\}
\end{equation}
%\vspace{-5}
\begin{equation}
\label{C2}
C_2= \mathbb{E} \left\{\log\det(\mathbf{I}_{M_d}+\mathbf{H}_{\textmd{MAC}}\left[
\begin{matrix}
\frac{P_s}{N_s}.\mathbf{I}_{N_s} &  \mathbf{0}\\
\mathbf{0}                                            &  \frac{P_r}{N_r}.\mathbf{I}_{N_r}
\end{matrix}
\right]\mathbf{H}_{\textmd{MAC}}^\dag)\right\}.
\end{equation}

\subsection{Decode-and-Forward Achievable Rate}
%\subsection*{\textbf{Decode-and-Forward Achievable Rate}}
The capacity of the full-duplex relay channel is lower bounded by the DF achievable rate \cite{cover1979capacity} given as
\begin{equation}
\label{eq13}
    R_{DF} = \max_{p(x_s,x_r)} \min{\{I(X_s;Y_r \mid X_r),I(X_s, X_r; Y_d)\}}
\end{equation}
where the optimal distribution is again Gaussian and the maximization should be done over the joint distribution of the source and relay signals. In this strategy the relay first decodes the received signal from the source, and then re-encodes it before forwarding it to the destination. Considering \eqref{pathloss.H} and using the same approach for evaluation of the CS upper bound, it is easy to show that the capacity of MIMO relay channel with only CSIR is lower bounded by
\begin{equation}
    \label{R.DF}
    R_{DF}=\min(C_{3},C_2)
\end{equation}
%\vspace{-10}
\begin{equation}
    \label{effecC3}
    C_3 = \mathbb{E} \left\{\log\det(\mathbf{I}_{M_r}+ \frac{P_s}{N_s}.{r_R}^{-\alpha}.\mathbf{\hat{H}}_{r,s}\mathbf{\hat{H}}_{r,s}^ \dag)\right\}
\end{equation}
\begin{footnotesize}
\begin{equation}
    \label{effecC2}
    C_2= \mathbb{E} \left\{\! \log\det(\mathbf{I}_{M_d}\!+\! \mathbf{\hat{H}}_{\textmd{MAC}}\! \left[
\begin{matrix}
\frac{P_s}{N_s}.\frac{1}{{r_D}^{\alpha}}.\mathbf{I}_{N_s} &  \mathbf{0}\\
\mathbf{0}                                            &  \frac{P_r}{N_r}.\frac{1}{{r_{DR}}^{\alpha}}.\mathbf{I}_{N_r}
\end{matrix}
\right] \! \mathbf{\hat
{H}}_{\textmd{MAC}}^\dag)\! \right\}.
\end{equation}
\end{footnotesize}

Note that \eqref{effecC3} and \eqref{effecC2} are the ergodic capacity of MIMO channel in source-relay and source-destination, respectively, and we will use them to find the optimal relay location in the sense of maximizing coverage region.

%%%%%%%%%%%%%%%%%%%%%%%%%%%%%%%%%%%%%
\section{Desired Transmission Rate Analysis}
\label{Sec.IV}
The optimal relay location $ d^* = r_R^*$ in the terms of desired transmission rate $R_c$ has been studied in \cite{alizadeh2012analysis} and \cite{alizadeh2013analysis}. In order to obtain a theoretical expression between the desired transmission rate $R_c$ and the optimal relay location $d^*$, the authors evaluated $R_c$ by using two analytical approaches. We recapitulate the main results of the high-SNR regime  in this section.
%\vspace{1}

In the high-SNR regime, the relation between the desired transmission rate $R_c$ and the optimal relay location $d^*$ can be approximated by the following expression
\begin{equation}
\label{eq22}
R\approx m.\log\left(\frac{ \rho .\exp^{\psi(1)}}{N_s}\right)+\frac{1}{\ln 2}.\sum^m_{p=1}\sum^{n-p}_{q=1}\frac{1}{q}
\end{equation}
where $-\psi(1)\approx 0.577215$ is the Euler-Mascheroni constant; $\rho = P_s .{d^*}^{-\alpha} $; $m = \min \left(N_s , M_r\right)$ and $n = \max \left(N_s , M_r\right)$.
%%%%%%%%%%%%%%%%%%%
% [width=3.5in]
\begin{figure}[t]
\centering
\includegraphics[scale=1]{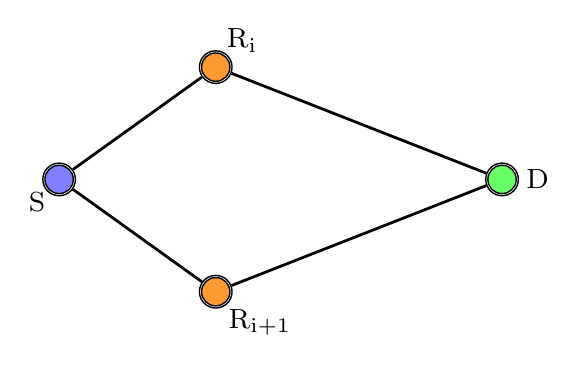}
    \caption{Relays' cooperation (Diamond Channel).}
    \label{Fig3}
\end{figure}
%%%%%%%%%%%%%%%%%%% 
%%%%%%%%%%%%%%%%%%%
%\begin{figure}[t]
%\centering
%\begin{tikzpicture}[scale=0.9]  % [width=\linewidth]
%\node at (2.5,0) {\footnotesize };
%\node at (2.1,-1) {\footnotesize S};
%\node at (7.9,-1) {\footnotesize D};
%\node at (2.8+1.7678,0.7678+0.3-0.5) {\footnotesize $\mathrm{R_i}$};
%\node at (3+1.7678,-2.7678-0.35+0.5) {\footnotesize $\mathrm{R_{i+1}}$};
%\draw [fill=blue!50] (2.5,-1) circle [radius=0.16];
%\draw [fill=green!60] (7.5,-1) circle [radius=0.16];
%\draw [fill=orange!80] (2.5+1.7678,0.7678-0.5) circle [radius=0.16];
%\draw [fill=orange!80] (2.5+1.7678,-2.7678+0.5) circle [radius=0.16];
%%%\draw (2.5,0) to [out=180,in=90] (1.5,-1) ;
%%%\draw [dotted] (1.5,-1) to [out=270,in=180] (2.5,-2);
%\node[circle,draw]
%(A) at (2.5,-1);
%\node[circle,draw]
%(B) at (7.5,-1);
%\node[circle,draw]
%(D) at (2.5+1.7678,0.7678-0.5);
%\node[circle,draw]
%(E) at (2.5+1.7678,-2.7678+0.5);
%\thicklines
%\draw[-,thick] (A) -- (D);
%\draw[-,thick] (A) -- (E);
%\draw[-,thick] (D) -- (B);
%\draw[-,thick] (E) -- (B);
%%%\draw[dotted,thick] (A) -- (F);
%%%\draw[dotted,thick] (B) -- (F);
%\end{tikzpicture}
%    \caption{Relays' cooperation (Diamond Channel).}
%	\label{Fig3}
%\end{figure}
%%%%%%%%%%%%%%%%%%% 
%\section{Cooperative Relays}
\section{Main Results}
In this section we study the situation in which two neighboring relays cooperate and the destination node receives signals from the two nearest relays which is known as the Diamond channel. Our channel model is depicted in Fig. 3. The received signal can be written as
\begin{align}
    \mathbf{y}_d &= \mathbf{H}_{d,s}\mathbf{x}_s + \mathbf{H}_{d,r_j}\mathbf{x}_{r_j} + \mathbf{H}_{d,r_{j+1}}\mathbf{x}_{r_{j+1}} + \textbf{z}_d
\end{align}

Intuitively, when received power at the destination is increased, then the sum-rate is increased. So, for a lower sum-rate (lower required average received power), the coverage region will be increased. In order to obtain a theoretical expression for coverage region extension factor, in what follows, we evaluate extension factor by using two  analytical approaches.
Assuming noncooperative scenario, the sum-rate capacity of the MIMO MAC is given by \cite{biglieri2007mimo}
\begin{IEEEeqnarray}{rcl}
R_{R}\! +\!R_{S} &<&\mathbb{E}_\mathbf{H}\!\left\{\! \log \det \left(\!\mathbf{I}_{M_d}\!\!+\!\!\frac{\rho_{ds}}{N_s} \mathbf{H}_{d,s}  \mathbf{H}_{d,s}^\dag \! +\! \frac{\rho_{dr}}{N_r} \mathbf{H}_{d,r} \mathbf{H}_{d,r}^\dag \! \right)\right\}\nonumber \\
&=&  \mathcal{C}_{\textmd{MAC}}^{\textmd{sum}}
\end{IEEEeqnarray}
where $\rho_{ds} = P_s.r_D^{-\alpha}$ and $\rho_{dr} = P_r.r_{DR}^{-\alpha}$; and $\mathcal{C}_{\textmd{MAC}}^{\textmd{sum}}$ denotes for the upper bound of MIMO MAC sum-rate.

\subsection{High-SNR Approximation}
%\subsection*{\textbf{High-SNR Approximation}}
Assume the destination is placed at the same distance from the source and the relay, also $N_s = N_r$ and $P_s = P_r$. For the best case, we have $\mathbf{H}_{d,s}  \mathbf{H}_{d,s}^\dag \approx \mathbf{H}_{d,r} \mathbf{H}_{d,r}^\dag$. Therefore, we can write the sum-rate as a function of the singular values, $\lambda _i $, of the random channel matrix $\mathbf{H}_{d,r}$. By Jensen's inequality, we get \cite{wirelessTse}
\begin{align}
\mathcal{C}_{\textmd{MAC}}^{\textmd{sum}} \leqslant r \log\left(1+\frac{2 \rho_{dr}}{N_r} \left[\frac{1}{r} \sum_{i=1}^r \lambda _i^2 \right]\right)
\end{align}
where $r = \min \{N_r , M_d \}$ is the rank of matrix $\mathbf{H}_{d,r}$ and $\lambda_1 \geq \lambda_2 \geq ...  \geq \lambda_r $ are the ordered singular values of  $\mathbf{H}_{d,r}$. At high SNR, we get
\begin{align}
\mathcal{C}_{\textmd{MAC}}^{\textmd{sum}} \approx  r \log \frac{\rho}{N_r}+ \sum _{i =1}^{r} \mathbb{E} \left\{\log \lambda _i^2\right\}
\end{align}
where $\rho = 2  \rho_{dr}$. Also, we have
\begin{align}
\sum _{i =1}^{r} \mathbb{E} \left\{\log \lambda _i^2\right\} = \sum _{i =\mid N_r - M_d \mid +1}^{\max \left\{N_r , M_d \right\}} \mathbb{E} \left\{\log \chi _{2i}^2\right\}
\end{align}
where $\chi _{2i}^2$ is chi-square distribution with $2i$ degrees of freedom.

\subsection{Low-SNR Approximation}
%\subsection*{\textbf{Low-SNR Approximation}}
Assume the destination is placed at borders of the coverage region, so ${\Vert \mathbf{H}_{d,s} \Vert}_F^2  \ll {\Vert \mathbf{H}_{d,r} \Vert} _F^2 $. We can ignore, therefore, the signal from the source and get the sum-rate as
\begin{align}
\mathcal{C}_{\textmd{MAC}}^{\textmd{sum}} &\approx \sum _{i =1}^{r} \frac{\rho_{dr}}{N_r} \mathbb{E} \left\{\log \lambda _{i}^2\right\} \log _2 e \nonumber \\
&=\frac{\rho_{dr}}{N_r} \mathbb{E} \left\{\mathrm{tr}\left[\mathbf{H}_{d,r} \mathbf{H}_{d,r}^\dag \right]\right\} \log _2 e \nonumber \\
&= \frac{\rho_{dr}}{N_r} \mathbb{E} \left\{\sum_i \sum_j \mid h_{ij} \mid ^2\right\} \log _2 e \nonumber \\
&= M_d ~\rho_{dr} \log _2 e \approx M_d ~ log _2 \left(1 + \rho_{dr} \right).
\end{align}

Generally, we can approximate sum-rate as
\begin{align} \label{sumerate.aprox}
    \mathcal{C}_{\textmd{MAC}}^{\textmd{sum}} \left(P_d\right) \approx K_1 \log  (1 + K_2 P_d)
\end{align}
where $P_d = P_{d,r}+ P_{d,s}$ is the received power at the destination. Assume that the destination receives signals from another nearest relay too (cooperative scenario), and $\hat{P}_d$ denotes for the required average received power needed at the destination in noncooperative scenario to achieve the same sum-rate as the cooperative scenario. Therefore, we get
\begin{align}
    \mathcal{C}_{\textmd{MAC}}^{\textmd{sum}} \left(\hat{P}_d\right) = \gamma \;  \mathcal{C}_{\textmd{MAC}}^{\textmd{sum}} \left(P_d\right)
\end{align}
where $\gamma \geq 1$ is the ratio of cooperative sum-rate to noncooperative sum-rate. Combining this with (\ref{sumerate.aprox}), results in
\begin{align} \label{power.ratio}
    \frac{P_d}{\hat{P}_d} = \frac{K_2 P_d}{\left(1 + K_2 P_d\right)^\gamma - 1}\, .
\end{align}

Going back to our model, we assume that the destination is at a distance $r_D$ from the source, and $r_{DR}$ from the relay which lies in its corresponding sector. In previous sections, we let $P_s$ and $P_r$ denote the average power transmitted by the source and relay, respectively; and $P_d$ denotes the average received power at the destination. Then, we have
\begin{align}
    P_{d_{\mathrm{(dB)}}} &= \left(P_{s_{\mathrm{\left(dB\right)}}} - PL(r_D)\right) + \left(P_{r_{\mathrm{(dB)}}} - PL(r_{DR})\right)\nonumber \\
    PL(d) &\triangleq PL(r_D)+PL(r_{DR}) = P_{s_{\mathrm{(dB)}}} + P_{r_{\mathrm{(dB)}}} - P_{d_{\mathrm{(dB)}}} \nonumber
\end{align}
where $d$ is a function of $r_D$ and $r_{DR}$. We use Hata formula for propagation loss from \cite{hata1980empirical} which is modeled as
\begin{align}
    PL_{\mathrm{(dB)}}(d) = A + B \log_{10} d
\end{align}
where $A$ and $B$ are frequency and transceiver antenna height functions. Hence,% and $d$ is the distance. Therefore
\begin{align}
    %PL(d) = PL(r_D) + PL(r_D^{'}) = P_S + P_R - P_D. \nonumber
    \log_{10} d = \frac{1}{B} \left(P_{s_{\mathrm{(dB)}}} + P_{r_{\mathrm{(dB)}}} - P_{d_{\mathrm{(dB)}}} - A\right).
\end{align}

We intend to rewrite this expression as a function of average received power at the destination and maximum distance $d_{\max}$. First, we assume $P_{d_{\mathrm{(dB)}}}$ is fixed and so the maximum distance $d_{\max}$ corresponds to the maximum transmitted power $\left(P_s + P_r\right)_{\max}$. We let $P_{\max}^T$ denotes for the maximum power that the source and relay can transmit, in decibels. We get
\begin{align}
    \log_{10} d_{\max}\left(P_{d_{\mathrm{(dB)}}}\right) = \frac{1}{B} \left(P_{\max}^T - P_{d_{\mathrm{(dB)}}} - A\right).
\end{align}

%%%%%%%%%%%%%%%%%%%%%%
\begin{figure}%[!t]
    \centering
    \includegraphics[width=3.55in]{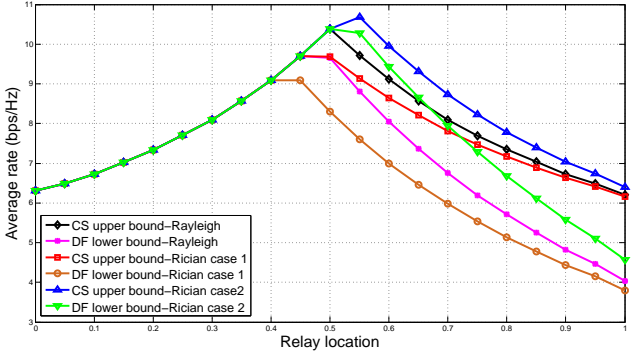}
    \caption{Comparison of capacity bounds for $P_s=P_r=10$ dB, $d_y = 0.1$, and $N_a=2$}
    \label{Fig4}
\end{figure}
%%%%%%%%%%%%%%%%%%%%%%%

Since the maximum transmitted power of transmitters (the source and relay) are fixed, and parameters $A$ and $B$ are fixed in a typical environment, then $P_{d_{\mathrm{(dB)}}}$ is the only parameter that determines the maximum distance $d_{\max}$. Furthermore, there is an inverse relation between maximum coverage and desired average received power.

Now, we compare the maximum coverage distance as a function of $P_d$ and $\hat{P}_d$
\begin{align}
    \log_{10} \hat{d}_{\max}(\! \hat{P}_{d_{\mathrm{(dB)}}}\!) \!- \! \log_{10} d_{\max}(\! P_{d_{\mathrm{(dB)}}}\! ) \! = \! \frac{1}{B} \left(\! P_{d_{\mathrm{(dB)}}} \! - \!\hat{P}_{d_{\mathrm{(dB)}}} \! \right). \nonumber
\end{align}
This yields
%\vspace{-3pt}
\begin{align}
     \frac{\hat{d}_{\max}(\hat{P}_{d_{\mathrm{(dB)}}})}{d_{\max}(P_{d_{\mathrm{(dB)}}})} = \left(\frac{P_d}{\hat{P}_d}\right)^{\sfrac{1}{B}}.
\end{align}
Combining this with (\ref{power.ratio}), results in the following expression for the coverage region extension factor:
\begin{align}
     \frac{\hat{d}_{\max}(\hat{P}_{d_{\mathrm{(dB)}}})}{d_{\max}(P_{d_{\mathrm{(dB)}}})} = \left(\frac{K_2 P_d}{\left(1 + K_2 P_d\right)^\gamma - 1}\right)^{\sfrac{1}{B}}.
\end{align}

%%%%%%%%%%%%%%%%%%%%%%%%%%%%%%%%%%%%%%%
\section{Simulation Results}

In this section, we consider three different fading models for source-relay channel ($\mathbf{H}_{r,s}$) and study the effect of correlation and LOS components of channel matrices on the capacity bounds and coverage region of MIMO relay network. In our simulations, we assume that $\alpha=3.52$, $P_s=P_r$, and all transmitters and receivers are equipped with two antennas, i.e., $N_s=M_r=N_r=M_d=N_a=2$.

To study the influence of the LOS component on the coverage region of MIMO relay network, ignoring phase factors, we consider two different LOS components for source-relay channel as follows \cite{paulraj2003introduction}
%\mathbf{\widetilde{H}}_{r,s}^{\textrm{LOS1}}
\begin{align}
\label{losmatrices}
%\mathbf{H}^{'2}_{2 \times 2} =
%\mathbf{\widetilde{H}}_{1}^{\textrm{LOS}}=
\mathbf{H}_{1}^{\textrm{LOS}}=
\begin{bmatrix}
1& 1\\
1& 1
\end{bmatrix}, \qquad
%\end{align}
%\begin{align}
%\mathbf{H}^{'2}_{2 \times 2} =
%\mathbf{\widetilde{H}}_{2}^{\textrm{LOS}}=
\mathbf{H}_{2}^{\textrm{LOS}}=
\begin{bmatrix}
1& -1\\
1& 1
\end{bmatrix}
\end{align}
where $\mathbf{H}_{1}^{\textrm{LOS}}$ corresponds to poorly-conditioned channel (Rician fading case 1) and $\mathbf{H}_{2}^{\textrm{LOS}}$ corresponds to well-conditioned channel (Rician fading case 2). The first case occurs when $r_D$ (source-relay distance) is much greater than the element separation at source antennas, whereas the second case generally occurs when $r_D$ is comparable to the element separation at the source or relay antennas.
%%%%%%%%%%%%%%%%%%%%%%%
\begin{figure}%[!t]
    \centering
    \includegraphics[width=3.55in]{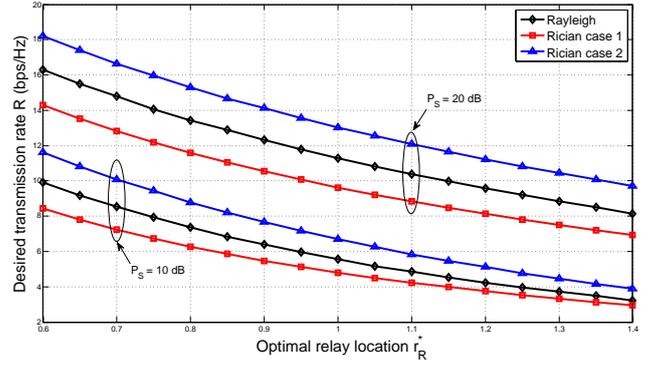}
    \caption{The relation between desired transmission rate and optimal relay location for $N_a=2$}
    \label{Fig5}
\end{figure}
%%%%%%%%%%%%%%%%%%%%%%% 

Considering the single relay case, we investigate the effect of fading on the capacity bounds, optimal relay location, and coverage region of our specific MIMO relay channel. Fig.~\ref{Fig4} depicts the effect of channel fading on the capacity bounds of MIMO relay channel. In this figure, we assume that the source and destination are located at $(0,0)$ and $(1,0)$, respectively, and the relay is located at $(d_x,d_y)$; all in Cartesian coordinates, where $d_y=0.1$ and $d_x$ is changing from 0 to 1. It can be observed that the source-relay channel fading has a significant effect on the capacity bounds. As it is shown in Fig.~\ref{Fig4}, since the second channel is orthogonal (rank $(\mathbf{H}^{\textrm{LOS}}_2) = 2$), while the first channel is rank-deficient (rank $(\mathbf{H}^{\textrm{LOS}}_1) = 1$) ,  the $\mathbf{H}^{\textrm{LOS}}_2$ channel outperforms the $\mathbf{H}^{\textrm{LOS}}_1$ channel. However, because perfect orthogonality of $\mathbf{H}^{\textrm{LOS}}_2$ requires specific antenna location and geometry, the first case is more possible in LOS propagation environment.
 
%%%%%%%%%%%%%%%%%%%%
Fig.~\ref{Fig5} depicts the desired transmission rate for different values of optimal relay location $r_R^*$ and compares the optimal relay location $r_R^*$ for two different values of transmit power $P_{s}$. From this figure, it is possible to find the boundary distance for applying DF strategy and determine the optimal relay location maximizing the coverage region. As it can be seen, the region below each curve represents the region where DF strategy can be applied, while the region above contains the points in which this strategy can not be used anymore.
It can be inferred from the figure that for a fixed desired transmission rate, the optimal relay location of Rician fading case 2 is larger than the other two cases; and since the coverage region has a straight relation with the optimal relay location, as we expect, this fading model outperforms the other two models in terms of coverage region.

%%%%%%%%%%%%%%%%%%%
\begin{figure}%[!t]
    \centering
    \includegraphics[width=2.96in]{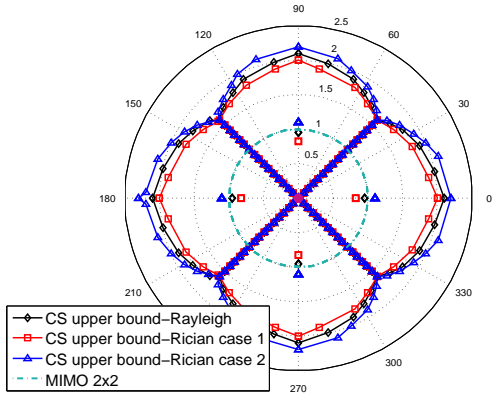}
    \caption{Comparison of coverage region for different fading channels with \!$L\! =\! 4$}
    \label{Fig6}
\end{figure}

%%%%%%%%%%%%%%%%%%%%
%%%%%%%%%%%%%%%%%%%%
\begin{figure}%[!t]
\centering
\includegraphics[width=2.97in]{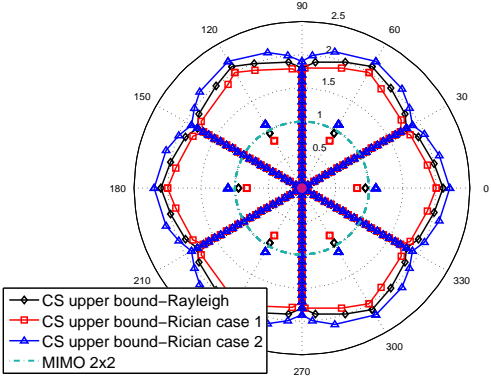}
 %where an .eps filename suffix will be assumed under latex,
% and a .pdf suffix will be assumed for pdflatex; or what has been declared
% via \DeclareGraphicsExtensions.
\caption{Comparison of coverage region for different fading channels with $L = 6$}
\label{Fig7}
\end{figure}
%%%%%%%%%%%%%%%%%%%%
%\vspace{-6pt}
\subsection{Influence of Fading on the Coverage Region}
In the following simulations, assuming $P_s=P_r=10$ dB, we use polar coordinates to illustrate the influence of fading on the coverage region of MIMO relay network. Considering the desired transmission rate $R_c=5.5$ bps/Hz, the respective optimal relay locations $r^*_R$ for Rayleigh fading, Rician fading case 1, and Rician fading case 2 are 1, 0.87, and 1.15. As stated before, these values are boundary distances which determine whether the DF strategy can be applied or not; thus, in order to guarantee that the relay is still able to decode the transmitted signal from the source, it is necessary to place relays at slightly shorter distances, i.e., 0.95, 0.82, and 1.1. Fig.~\ref{Fig6}~and ~Fig.~\ref{Fig7} evaluate the fading effect of source-relay channel on the coverage region of our network. It is clear from the figure that the Rician fading case 2 provides a wider coverage region.
%\vspace{-5pt}
\subsection{Influence of Cooperation on the Coverage Region}
\begin{figure}
\includegraphics[scale=0.92]{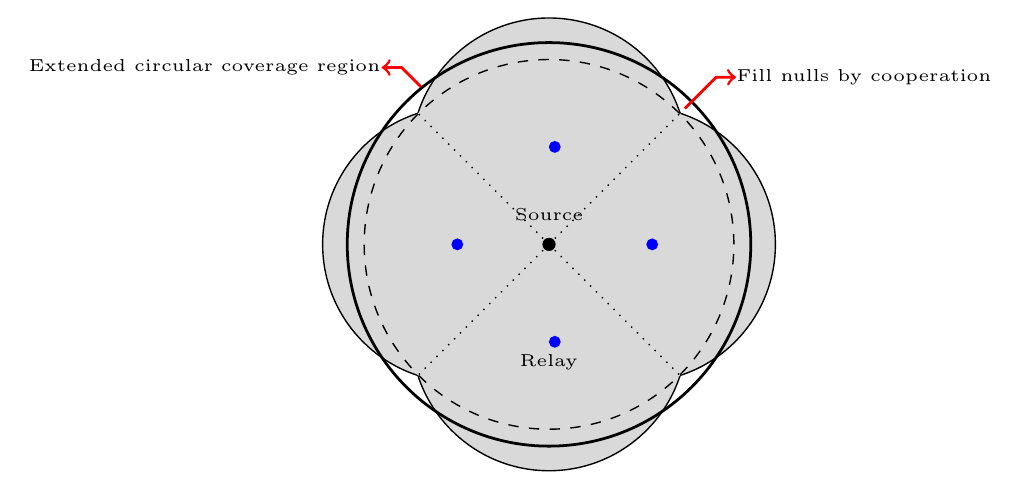}
\caption{Extension of circular coverage region in cooperative scenario}
\label{Fig8}
\end{figure}

%\textcolor{red}{!!!! the most important result of the paper !!!!}
Fig.~\ref{Fig8} show the network with four relays. For performance comparison, this figure include the circular coverage region of the cooperative scenario where each two neighboring relays cooperate in transmission. As it is shown, the circular coverage region enhancement is obvious, particularly in nulls.

%\vspace{-5pt}
\section{Conclusion}
In this paper, we analyzed coverage region for MIMO relay network consisting of $L$ relays which are located uniformly on a circle around the source. Considering three fading models, coverage region, capacity bounds and optimal relay location were obtained for this models. 
%Also, we showed that increasing the number of relays enlarges the circular coverage region. 
Finally, we studied the situation in which the two adjacent relays cooperate with the destination. In this case, as it is shown by our main results, the circular coverage region is increased.

\bibliographystyle{IEEEtran}
%\bibliography{references}
\bibliography{ref}
% Press F11 and then F1 to apply any change in references.bib

%\begin{thebibliography}{1}


%\end{thebibliography}

\end{document}